\documentclass{ws-jai}
\usepackage[flushleft]{threeparttable}

\usepackage[utf8]{inputenc}

\usepackage{hyperref}
\usepackage{aas_macros}
\usepackage{amsmath}
\usepackage{comment}

\begin{document}

\markboth{Weiwei Chen}{Wide field Beamformed Observation with MeerKAT}

\title{Wide field Beamformed Observation with MeerKAT}

\author{Weiwei Chen$^{1}$, Ewan Barr$^{1}$, Ramesh Karuppusamy$^{1}$, Michael Kramer$^{1,2}$, Benjamin Stappers$^{2}$}

\address{
$^1$ Max-Planck-Institut f\"{u}r Radioastronomie, Auf dem H\"{u}gel 69, D-53121 Bonn, Germany\\
$^2$ Jodrell Bank Centre for Astrophysics, University of Manchester, M13 9PL, UK
}

\maketitle

\begin{abstract}
    Large-scale beamforming with radio interferometers has the potential to revolutionize the science done with pulsars and fast radio bursts by improving the survey efficiency for these sources. We describe a wide-field beamformer for the MeerKAT radio telescope and outline strategies to optimally design such surveys. A software implementation of these techniques, \textsc{Mosaic} is introduced and its application in the MeerKAT telescope is presented. We show initial results using the beamformer by observing a globular cluster to track several pulsars simultaneously and demonstrate the source localization capability of this observation.   
\end{abstract}


\section{Introduction}

Radio pulsars are now accepted as spinning neutron stars and have become important tools for testing fundamental physics. For instance, the binary pulsar system J0737-3030A/B serves as an excellent natural laboratory for testing theories of gravity \citep{Kramer2004} and the discovery of massive pulsars has constrained models of the density of nuclear matter \citep{Dem2010, Ant2013}. Recently, the discovery of Fast Radio Bursts (FRBs) has provided new tools for the study of the Universe at cosmic distances \citep{Lorimer2007} and the measurement of the cosmic baryonic density \citep{mpm+2020}. While over 500 FRBs have now been discovered \citep{pbj+2016, The_CHIME_FRB_Collaboration2021}, the nature of the progenitors is unclear. An increase in the number of FRBs and pulsars, or the discovery of new unknown transient phenomena, motivates the continuation of time domain searches in signals of astronomical origins.

\par Traditionally, large single dish telescopes have been employed in pulsar and transient searches, despite their small fields of view (FoV) being a major drawback in large surveys.  Often, multi-beam receivers or phased array feeds (PAFs) augment single-dish telescopes to increase the FoV, thereby increasing the survey speed \citep{smits2009}. For instance, the 13-beam receiver on the Parkes radio telescope \citep{Staveley-Smith1996}, the 7-beam on the Arecibo Telescope \citep{Cordes2006} and 19-beam receiver on FAST \cite{Han2021} have increased the FoV by a factor of 10 and a similar increase in survey speed when compared to single-beam receivers. Other than the increase in FoV, precise localisation is key to the unambiguous association of the FRBs to their host galaxy \citep[see e.g.][]{mpm+2020}. Similarly, good initial localisation of newly discovered pulsars greatly reduces the time required to converge on an accurate timing model, with which tests of fundamental physics can be done. Lastly, multi-beam observations have become highly desirable for the best FRB and pulsar searches, as they allow for robust discrimination against terrestrially generated RFI.

\par The seemingly non-convergent requirements of large FoV and a relatively high spatial resolution at sensitivities rivaling large single-dishes are best addressed by radio interferometers comprised of many small dishes. Digital technology has also evolved to allow the generation of thousands of synthetic beams on sky with which the instantaneous FoV can be increased. However, the interferometrically synthesised beam is more complex than its single dish counterpart. Beam shape and the subtleties thereof are also crucial to any source localisation that follows, as evident in the initial localisation of the first repeating FRB \citep{Spitler2014}. This problem is exacerbated in the synthesised beam from an interferometer; the beam shape depends on time and the telescope altitude-azimuth positions. When several such beams are used to tile a given region in sky,  the resulting FoV changes continually over the course of an astronomical observation. These intricacies have to be fully characterised for any successful fast radio transient survey. To this end, we develop a method to model the synthesised beam shapes and the aforementioned FoV variation.

\par We apply these techniques to the MeerKAT interferometer \citep{Camilo2018} and present results of simulations and actual observations. MeerKAT is a 64-dish interferometer in the Northern Cape of South Africa, with baselines of up to $\sim$ 8$\,$km and a collecting area of $\sim$9161~m$^2$. This facility serves as a test bed for our beam generation and beam-packing methods. At the time of writing, the MeerKAT telescope provides one synthesized beam as an observatory facility. Through the use of an externally funded and developed total power beamformer \citep{Barr2018}, the FoV of beamformed observations can be improved by generating up to 1000 beams on the sky with high time and spatial resolution while achieving sensitivity on par with large single dishes. 

\par The remainder of this paper is organised as follows. We begin with some background on digital beamforming in \S \ref{sec:digital_beamforming}.  The beam shape, tiling technique to increase FoV and the change in FoV are discussed in the context of multi-beam observations in \S \ref{sec:multibeam_observations}.
Observations using such a beamformer must be carefully planned to account for the time, frequency and array configuration dependence of the synthesised beam and these aspects are discussed in \S \ref{sec:meerkat_obs} as a case study for the MeerKAT telescope. In \S \ref{sec:meerkat_obs}, we also present a demonstration of the beamformer that makes use of these techniques. We conclude by summarising this work in \S \ref{sec:summary}.

\section{Digital Beamforming}
\label{sec:digital_beamforming}

Beamforming modifies the spatial response of an array of sensors, resulting in an enhanced sensitivity in a particular direction. In the case of observations with a radio interferometer, this is done by the phase coherent summation of the voltages \citep[see e.g.][]{Van_Veen1998} from the individual receptors. The process can be expressed as  
\begin{equation}
F(\rm \theta, \phi) = \sum_{n=1}^{N} f(t_{\rm ref}-\tau_n)w(\tau_n).
\label{eq:beamforming}
\end{equation}
The result of the summation, $F(\rm \theta, \phi)$ in Equation \ref{eq:beamforming}, represents a directional beam at the desired pointing direction $\rm u(\theta, \phi)$, with a gain equivalent to the aggregate gain of $N$ antennas. $\rm t_{ref}$ is the time at the array reference, and $\tau_n$ is the difference between arrival times of wavefronts from $u(\theta, \phi)$ at antenna $n$ and the array reference. Coherence is achieved by referencing the voltages of antenna $n$, $f(t_{\rm ref}-\tau_n)$, to the array reference, through the application of delay correction $w(\tau_n)$. The delay $\tau_n$ is a function of $u(\theta, \phi)$ and the relative position of the antenna $r_n(x, y, z)$ in the array centered at the array reference as written in equation: 
\begin{equation}
\tau_n = \frac{u(\theta, \phi) \cdot r_n(x, y, z)}{c} 
\label{eq:voltage_beamforming}
\end{equation}
where $c$ is the speed of light. The delay $\tau_n$ only represents the geometric delay, any other delays in the system (e.g. instrumental delays) are not included. 

Beamforming can be implemented in multiple ways. Here we will only discuss digital beamforming and assume that an initial stage in which the ``coarse'' delay (the integer sample delay) and the ``fine'' delays (sub-sample delays) to the phase center are corrected for, leaving only the differential delays between the phase center and the directional beams remaining. These differential delays may be corrected for in either the time domain, through the application of a non-symmetric finite impulse response (FIR) filter \citep[see e.g.][]{frost1972}, or in the frequency domain following Fourier shift theorem. In the latter approach, the corrections may be applied directly to the voltages from the receptors or to the cross-correlations of those voltages, the so-called ``visibilities'' \cite[see e.g.][]{Roy2018}. \par 
When dealing directly with the voltages from the receptors, compensation for a delay $\tau$ in the time domain can be made by a phase shift of $e^{-j\omega\tau}$ in the Fourier domain:
\begin{equation}
\int_{-\infty}^{\infty} f(t-\tau)e^{-j\omega t} dt =  e^{-j\omega \tau} \int_{-\infty}^{\infty} f(t')e^{-j\omega t'} dt'\ ,   
\end{equation}
where $t' = t - \tau$. For broadband signals, it is first necessary to Fourier transform the voltages from each receptor such that the delays can then be compensated in the form of a frequency-dependent phase shift $e^{-j\omega\tau}$, often referred to as a ``weight''. While the approaches in the time domain and Fourier domain are analogous, delay compensation in the Fourier domain is used in the majority of modern radio interferometers as it allows for the reuse of the F--stage (Fourier transform) or X--stage (correlator) in an FX correlator. The choice of whether to use visibility-based or voltage-based beamforming is dependent on the science goals of an observation and the computational power available.  \par
As discussed in \citet{Roy2018}, to achieve the same FoV and time resolution with voltage and visibility beamforming, the visibility method has a higher instantaneous memory requirement than the voltage method by a factor of $\frac{N_{\rm ant} - 1}{2}$, where $N_{\rm ant}$ is the number of antennas. This is due to the fact that in the case of visibility beamforming one must integrate over all baselines in the array whereas in the case of voltage beamforming one must only integrate over all antennas in the array. The difference in computational cost between the two methods depends strongly on number of samples that are integrated in the correlator. For beamformers designed to produce high time resolution data (e.g. for instruments performing pulsar and fast transient searches) it is thus in general preferable to use voltage beamforming. If the time resolution required from the beamformer is such that $N_{\rm int} > \frac{N_{\rm ant} - 1}{2}$ then visibility beamforming is computationally cheaper. It should be noted that this only holds for arrays without redundant baselines. For arrays with redundant baselines, exploitation of the redundancy can reduce the cost of visibility beamforming. This paper focuses on applications that use voltage beamforming and that apply delay corrections via phase shifting in the Fourier domain.

\section{Multi-beam Observations}
\label{sec:multibeam_observations}
As noted in the introduction, multi-beam beamformers can greatly enhance the scientific capabilities of an interferometer through increasing point-source survey speed. However, to be effective in such surveys, care must be taken during observation planning as the survey speed and completeness is dependent on the ever-changing shape of the synthesised beams and the tiling of those beams on the sky. In this section, we discuss the simulation of the synthesised beam shape and the generation of efficient tiling patterns. \par
We can decompose the problem of efficient multi-beam observations into three components:
\begin{itemize}
  \item Determination of the synthesised beam shape
  \item Packing of the beams to generate a tiling
  \item Time-dependent evolution of the tiling pattern
\end{itemize} 

\par To solve this problem, we have developed \textsc{Mosaic}\footnote{\url{https://github.com/wchenastro/mosaic}}, a pure Python software package that implements tools to characterize the beam shape and generate efficient tilings. 

\subsection{Beam shape simulation}
\label{sec:beam_shape_sim}
Provided with an arbitrary set of antenna positions and a source position, the \textsc{Mosaic} package uses a sparse discrete Fourier transform approach to recover the point spread function (PSF, also known as the synthesized beam) of the array with a user-defined spatial resolution and FoV. The PSF is frequency dependent and the decision of which frequency to use for an observation is usually dictated by source and science goals. As such, frequency selection is handled on a case-by-case basis. By default, \textsc{Mosaic} weights each antenna in the array equally during the generation of the PSF. However, in real observations, the performance of the antennas and the quality of the calibration applied may vary. This results in non-uniform amplitude and phase responses across all antennas, causing the true PSF of the observation to differ from the simulation. To allow for more realistic PSF simulations, \textsc{Mosaic} accepts individual complex weights for each antenna. Manipulation of these weights allows the user to implement various apodization schemes \citep[see e.g.][]{Mort2016}. \par
\begin{figure}[h]
    \begin{center}
    \includegraphics[width=0.49\textwidth]{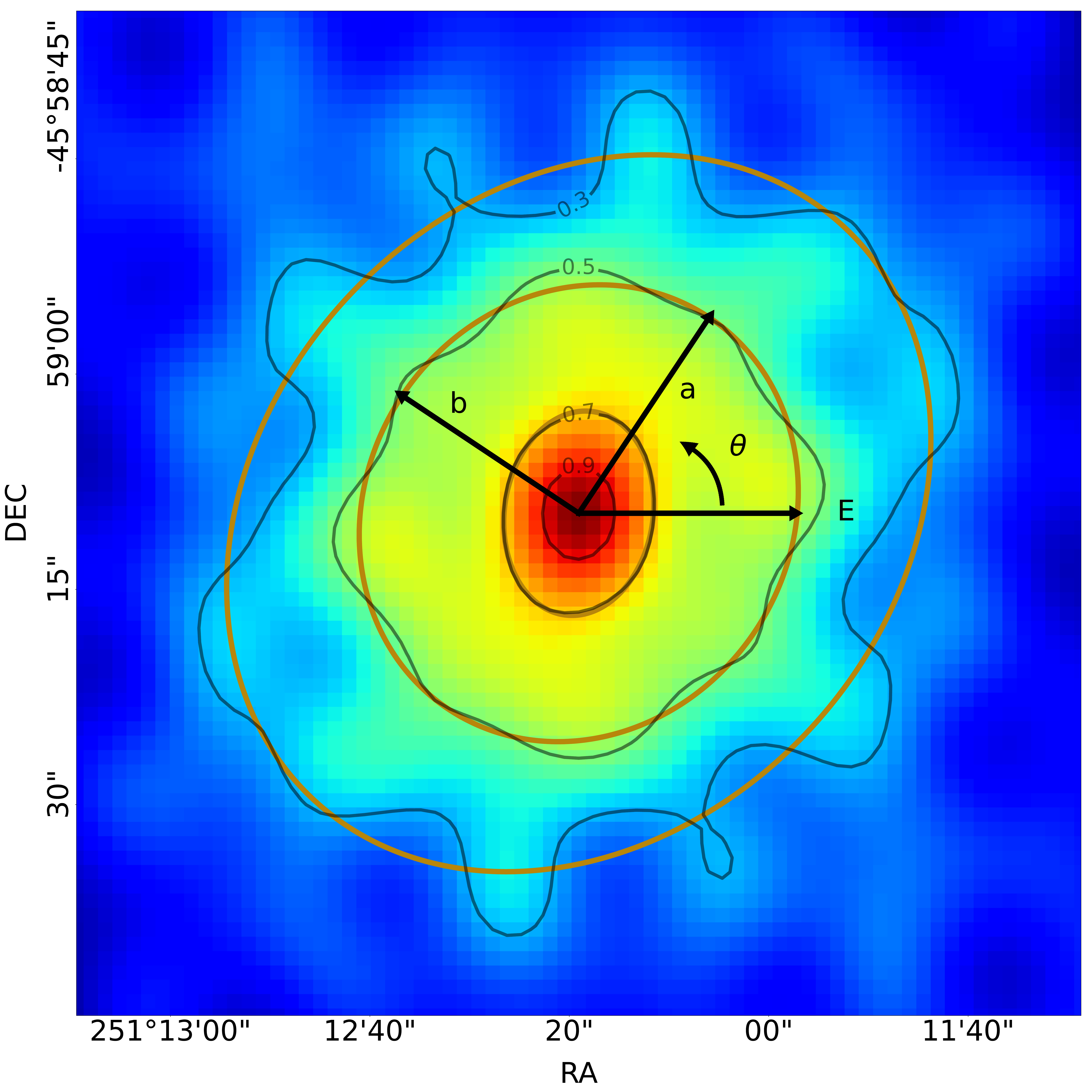}
    \end{center}
    \caption{Beam shape approximation: The background shows a simulated PSF for a MeerKAT observation using 54 equally weighted antennas. The orange ellipses represent the approximations of the beam at 30\%, 50\% and 70\% response levels as determined by the modelling procedure (see text). The semi-major, $a$, and semi-minor, $b$, axes and orientation angle, $\theta$,  of the modelled beam at the 50\% response level are also indicated. The orientation angle is measured from the east of the image plane. Black closed lines show the corresponding true contours of the PSF at the 30\%, 50\%, 70\% and 90\% response levels.}
    \label{fig:beamshape_fit}
\end{figure}

Due to the generally poor instantaneous UV coverage of interferometers, the shape of the PSF is often complex. As illustrated in Figure \ref{fig:beamshape_fit}, the region bounded by a certain response level may be irregular. However, for the purposes of later tessellation of beams in a tiling, it is helpful to approximate it as an ellipse. In this way, the beam shape is represented as a pair of semi-major, $a$, and semi-minor, $b$, axis lengths and the orientation of the beam, $\theta$, as shown in Figure \ref{fig:beamshape_fit}. However, a single set of semi-axes and angle neglects the variation of the beam shape at different response levels. For this reason, \textsc{Mosaic} produces a series of contours for the synthesized beam at different response levels. For each response level, the beam is approximated by fitting an ellipse to the respective contour, using an algorithm derived from \citet{Fitzgibbon1999}. In the end, this set of approximations is interpolated over to construct a smooth model of the beam shape as a function of response level that can be quickly looked-up during tiling.

\subsection{Beam tiling}
\begin{figure}[h!]
        \begin{center}
        \includegraphics[width=0.48\textwidth]{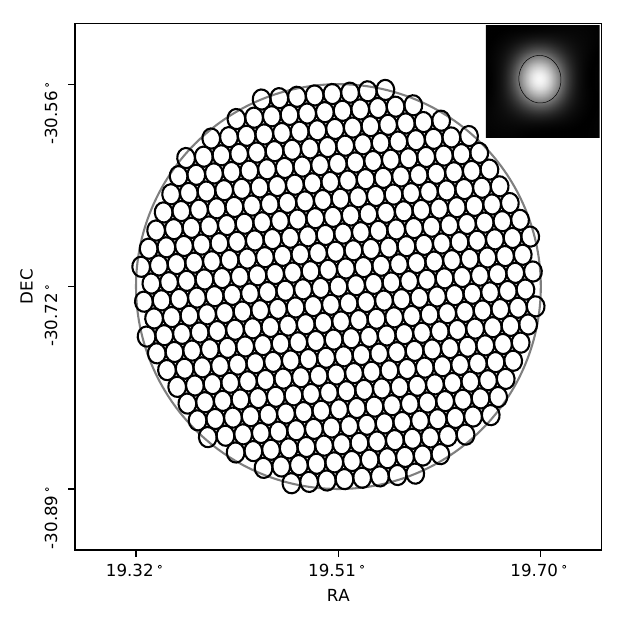}  \\
        \includegraphics[width=0.48\textwidth]{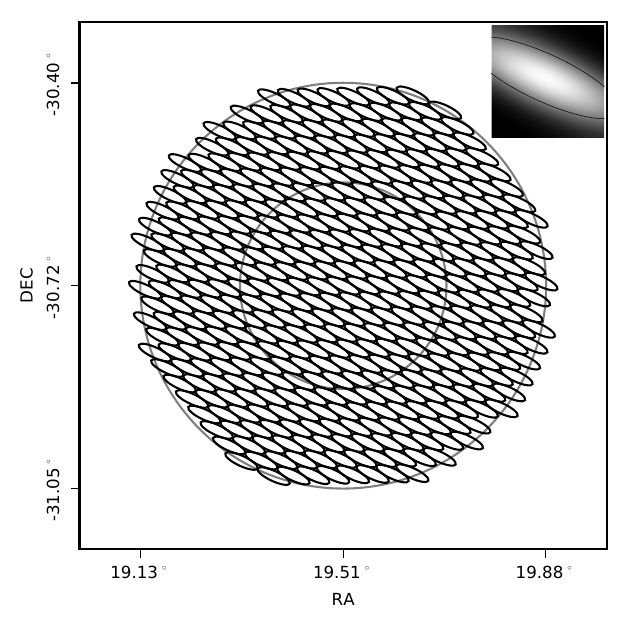}
        \end{center}
    \caption{Beam shape variety and tiling: The PSF, fitted beam shape and tiling configuration for the inner 44 antennas of the MeerKAT array when pointing at zenith (top, with 405 beams) and near the horizon (bottom, with 409 beams). The inner circle of the bottom panel is the size of the tiling of the top panel. The sky coverage of the tiling on the bottom is noticeably larger than the one on top due to the shallower projection angle of the array when observing at lower elevation.}
    \label{fig:beamshape_variety}
\end{figure}
With the beam shape approximated as a set of ellipses at different response levels, an efficient tessellation can be achieved through hexagonal packing. The sensitivity profile of the tiling is determined by the separation of the beams. Different science goals often require a different uniformity of sensitivity, as such the separation of the beams may be tuned by setting an ``overlap ratio'' parameter. This parameter determines the response level at which neighbouring beams intersect in the tiling pattern. For example, an overlap ratio of 50\% implies that the beams intersect at their fitted half-power points.\par
There are two methods to tile the beams depending on whether the user requires a fixed overlap (constant minimum sensitivity) or a fixed boundary of the tiling (constant FoV). 
In the case of fixed overlap, the tiling is created as follows:
\begin{enumerate}
    \item The tiling code retrieves the semi-axes and orientation from the beam shape model, according to the required overlap ratio. \label{itm:obtain_beamshape_fixed_overlap}
    \item The user selects either a circular (as seen in Figure \ref{fig:beamshape_variety}) or hexagonal bounding box and the tiling code calculates a possible radius or circumradius using the semi-axes and desired number of beams.\label{itm:choose_boundary}
    \item Twice this radius is used as a side length to create a square tiling of beams. Assuming the horizontal and vertical semi-axes of the beams are $a$ and $b$ (i.e. using an initial tiling frame in which $\theta=0$), to form a tessellation in the square tiling, beams in the same row are separated by $2a$ and neighbouring rows are vertically offset by $\sqrt{3}b$ and horizontally offset by $a$.  \label{itm:square_tiling}
    \item The tiling is rotated by the orientation angle, $\theta$, obtained from step \ref{itm:obtain_beamshape_fixed_overlap} such that tiled beams now have their true orientation in the image plane. \label{itm:rotate_tiling}
    \item Beams are filtered to retain only those which lie inside the bounding area chosen in step \ref{itm:choose_boundary}.
    \item If the absolute difference between the required number of beams and the number inside the bounding box is larger than a user-defined tolerance, the code will update the assumed bounding radius calculated in step \ref{itm:choose_boundary} and re-run the subsequent steps. This is repeated until the desired tolerance is achieved.
\end{enumerate}
If the user requires a fixed bounding box for the tiling, the following steps are used:
\begin{enumerate}
    \item The user specifies a bounding box in the form of a circle, hexagon, ellipse, arbitrary polygon or annulus. \label{itm:choose_fixed_boundary}
    \item The tiling module calculates a possible overlap ratio according to the scale of the bounding box and the desired number of beams. For the case of circles and hexagons, the area of the bounding box is used as a prior in this calculation, while for other shapes, the distance between the furthest point of the boundary and the tiling center is used as the prior. \label{itm:obtain_beamshape_fixed_size}
    \item The code retrieves the corresponding semi-axes and orientation angle from the beam model according to the overlap obtained from the step \ref{itm:obtain_beamshape_fixed_size}.
    \item It creates a square tiling and rotates it like in the steps \ref{itm:square_tiling} and \ref{itm:rotate_tiling} of the fixed overlap method, the only difference is that the side length is two times the distance between the furthest boundary point and the center of the tiling.
    \item The code filters the beams by which lie inside the specified boundary. For a polygonal boundary, the determination of whether a beam lies inside the boundary is done using an implementation\footnote{\url{https://wrf.ecse.rpi.edu/Research/Short_Notes/pnpoly.html}} of \citet{Shimrat1962}. For an annulus boundary, the beams inside the outer boundary and inner boundary are determined separately, with only beams which lie inside the outer boundary but not the inner boundary being retained. The annulus boundary is especially useful when combined with another tiling having the same size of the inner boundary of the annulus but with different overlap ratios as this allows for variable overlap ratios in a single pointing. 
    \item For this method, it is unlikely that the difference between the number of beams required and the number of beams inside the boundary is within the defined threshold in the first tiling attempt. The module tries to increase or decrease the overlap ratio to reduce the difference by performing an optimization routine with the overlap ratio as the free parameter. However, as discussed in section \ref{sec:beam_shape_sim}, the semi-axes and orientation can change with different overlap ratios. Therefore, it is necessary to obtain a new set of semi-axes and orientation angle whenever the overlap ratio is changed. This is also the reason for interpolating a smooth model in the last step of beam shape simulation so that the tiling module can obtain the beam shape in a specific overlap ratio quickly during the optimization routine without re-evaluating the PSF in every tiling attempt.
\end{enumerate}

Two examples of generated PSFs for MeerKAT (discussed further in Section \ref{sec:meerkat_obs}) and the corresponding circular tiling pattern are shown in Figure \ref{fig:beamshape_variety}. There are two important caveats to note here. First, we assume the desired tiling covers a sufficiently small solid angle, such that the shape of all beams in the tiling can be reasonably approximated by the  beam shape at the centre of the tiling. In the case of large tiling areas where non-isoplanactic effects become important, it is recommended that the tiling area be split into multiple sub-tilings such that each sub-tiling has a locally correct beam shape. Secondly, we assume that the synthesised beam may always be reasonably approximated by an ellipse at some response level. This assumption does not necessarily hold for all array configurations and response levels. Furthermore, the auto-optimized overlap ratio in a fixed boundary tiling is hard to predict due to the variable orientation angle of the beam at different response levels. As such, when a specific minimum sensitivity is required across the tiling pattern, it is encouraged to perform \textit{a priori} simulations of the tiling locally using \textsc{Mosaic} to ensure the result is acceptable.

\subsection{Tiling evolution}
\label{sec:tiling_evolution}

\begin{figure}[h]
    \begin{center}
    \includegraphics[width=0.5\textwidth]{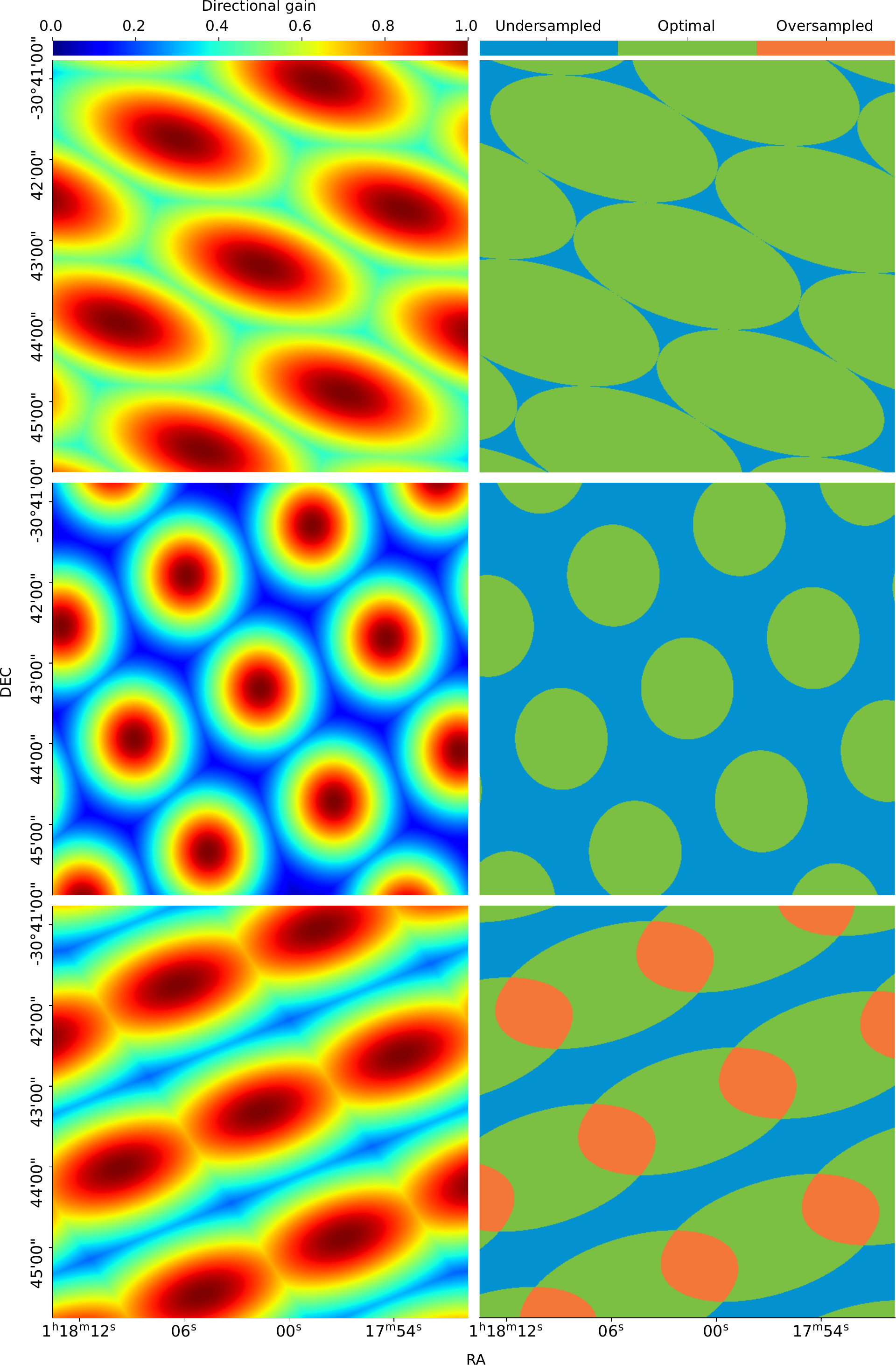}
    \end{center}
    \caption{Overlap evolution through time: The images on the left are the simulated direction gain of the array throughout the tiling. The images on the right are the profile of the beams with their widths plotted at their 50\% power level. The top row shows the situation at the start of the observation with an optimized tiling and an hour angle of 285$^{\circ}$. The middle row shows the situation 6 hours later with an hour angle of 15$^{\circ}$, large gaps exist between the beams. The bottom row shows the situation 10 hours after the start of the observation with an hour angle of 60$^{\circ}$ when the beams are overlapped with each other at a high ratio while large gaps exist at the same time. These simulations all have the same field of view.}
\label{fig:overlap_evolution}
\end{figure}

In the beam tiling process described above, the generated tiling is valid for a single array configuration, frequency and epoch. When tracking positions that are not in a fixed frame with respect to the array, there are two important effects that alter the tiling. The first is the array projection, whereby the instantaneous UV coverage of the array changes with the position of the source, thus changing the PSF and therefore also the approximated beam shape. The second is parallactic angle rotation whereby beams with fixed celestial coordinates will orbit the array boresight due to the rotation of the Earth. These effects are shown in Figure \ref{fig:overlap_evolution}. This changing of both beam shape and position results in tilings becoming inefficient over time as the beams become more or less overlapped and the sampling of the desired tiling region becomes less uniform. The magnitude of this effect can be simulated. However, the metric by which one evaluates the efficiency of the tiling is necessarily dependent on the goal of the observation. For example, a targeted point source observation with a relatively small extent may oversample the search region to allow for longer observations that are not as sensitive to the changing beam shape and orientation. Alternatively, a blind point source survey may optimise for survey speed by defining a maximum observation duration during which the tiling can remain above some efficiency threshold. For the purpose of demonstrating the effects of tiling evolution, in this article, we define tiling efficiency as the fraction of the tiling area which is neither oversampled nor undersampled. This idea is illustrated in the images on the right of Figure \ref{fig:overlap_evolution}. Here we treat all beams as their approximated ellipses and sample the tiling pattern to find regions covered by only one beam (shown in green), more than one beam (a proxy for excess overlap ratio or oversampling, shown in red), no beams (a proxy for undersampling, shown in blue).  \par
In practice, it is incumbent on the observer to simulate the evolution of the tiling efficiency over the course of an observation to make decisions on if and when re-tilings should occur. Re-tilings result in all beams changing their tracking positions and thus may be problematic for observation types that require unbroken integrations on specific positions, such as coherent pulsar searches. The rate of change of the tiling efficiency and its impact on observations is explored in greater depth in Section \ref{sec:tiling_validity}. The techniques described above are implemented in \textsc{Mosaic} for arbitrary arrays. Below we look specifically at beamforming with the MeerKAT array and discuss \textsc{Mosaic}'s deployment as part of the multi-beam beamformer.

\section{Multi-beam observations with MeerKAT}
\label{sec:meerkat_obs}

\subsection{The FBFUSE beamformer}

To support time domain pulsar and fast transient discovery at MeerKAT, the Max--Planck--Institute for Radio Astronomy has developed a multi-beam beamformer capable of forming of the order of 1000 total power beams using the full MeerKAT array. This beamformer, known as FBFUSE \citep[Filterbanking Beamformer User Supplied Equipment,][]{Barr2018}, provides the data products required by the TRAPUM experiment \citep{Stappers2016} to perform target searches for new pulsars, by the MeerTRAP project \citep{Sanidas2018} to perform real-time commensal searches for fast radio transients and by the MeerKAT Galactic Plane Survey (MGPS) to perform a wide-field survey for pulsars in the Galactic plane at L-band and S-band \cite{Kramer2016}. The technical details of the FBFUSE instrument will be addressed elsewhere. Here we will focus only on how the \textsc{Mosaic} software is used as part of FBFUSE and the implications of the effects described in Section \ref{sec:multibeam_observations} on observations. \par
\begin{figure*}[h!]
    \centerline{
    \includegraphics[width=0.9\textwidth]{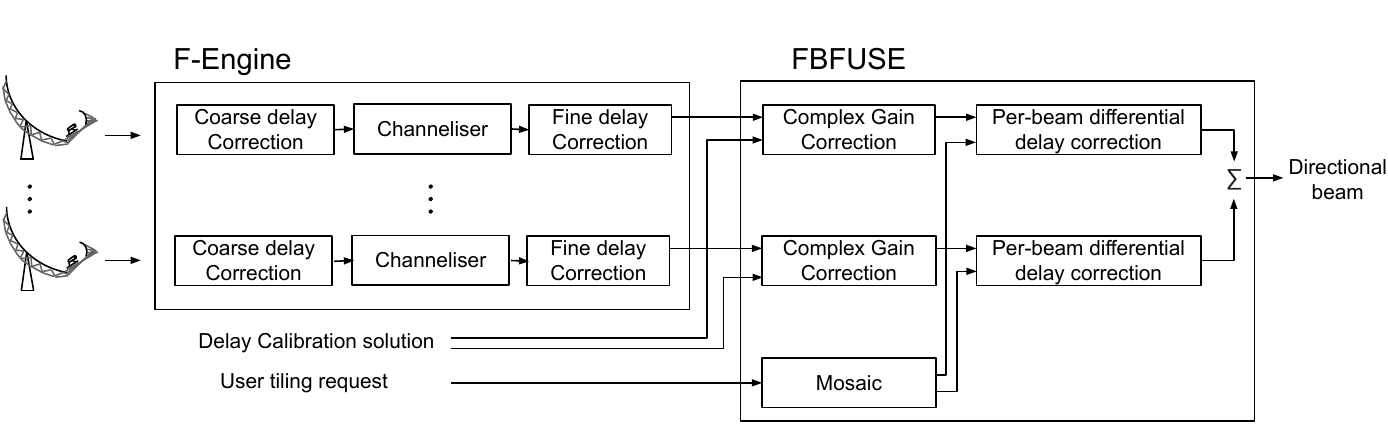}}
    \caption{Delay compensation chain for FBFUSE observations: The voltages from individual antennas first go through the F-engine where they are coarse delay corrected, channelized using a polyphase filterbank and fine delay corrected. After the F-engine, the voltages are all phased with reference to the telescope phase centre (usually the boresight pointing position). In FBFUSE a per-beam differential delay compensation calculated by $\textsc{Mosaic}$ is applied to the voltage to form coherent beams at sky positions defined by the tilings being used.}
    \label{fig:beamforming_data_flow}
\end{figure*}

The delay compensation chain for observations with FBFUSE is shown in Figure \ref{fig:beamforming_data_flow}. Voltages sampled at individual antennas are first propagated to the MeerKAT F-engines where channelisation is performed. The F-engines compensate for the geometric delay to the source under observation through first a coarse delay correction prior to channelisation and then a fine phase correction post channelisation. The channelised voltages from the F-engines are then transmitted via an Ethernet network to the FBFUSE instrument, where complex gain correction provided by the MeerKAT Science Processor sub-system is applied to correct for the instrumental delays among the streams from all antennas at all frequencies (depending on the telescope and observation configuration, this correction can also be applied inside the F-engines). At this stage, a direct summation of the voltages from each antenna would result in a coherent beam at the telescope phase centre. To produce tilings of off-boresight beams, FBFUSE invokes \textsc{Mosaic} to generate both an efficient tiling of beams on the sky and the corresponding delay polynomials required to re-phase the channelised voltage data to each beam position. FBFUSE uses calibration solutions provided by MeerKAT's Science Data processing (SDP) system \citep{Jonas2016}. \par
{}
\begin{figure}[h!]
   \centerline{
   \includegraphics[width=0.5\textwidth]{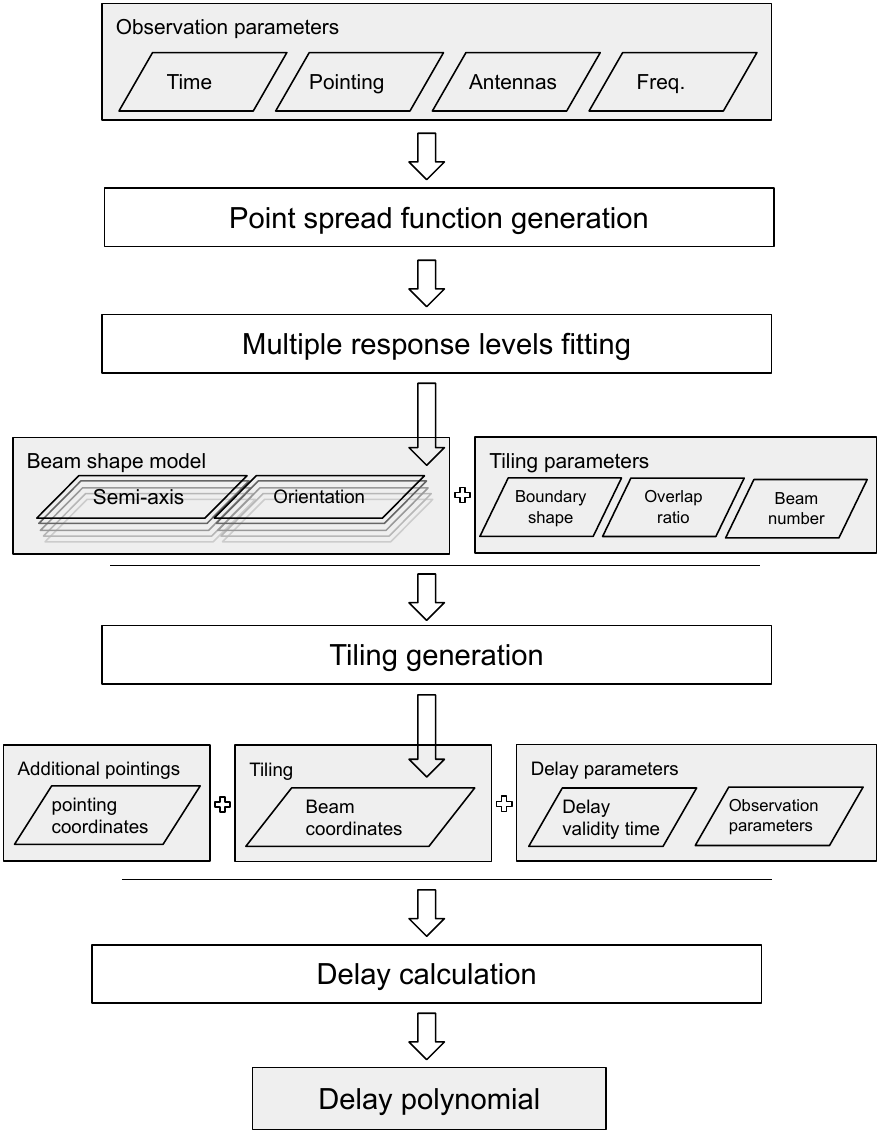}}
   \caption{Beamforming procedure. Process units are in white boxes and the grey boxes are inputs or outputs to or from the process units. The beam shape model is a table of semi-axis and angles indexed by response level.}
   \label{fig:beamforming_procedure}
\end{figure} \par
The tiling and delay pipeline used in \textsc{Mosaic} is shown in Figure \ref{fig:beamforming_procedure} and is described below: 
\begin{itemize}
  \item \textbf{Beam shape simulation:} FBFUSE retrieves the observer requests and provides \textsc{Mosaic} with the desired epoch of observation, the current MeerKAT array configuration and the observing frequency. Using these, \textsc{Mosaic} simulates the PSF at the array phase centre accordingly. A beam model is constructed by fitting ellipses to the contours of the PSF at different response levels. 
  \item \textbf{Tiling generation:} FBFUSE retrieves one or more predefined tiling configurations requested by the observer which include the number of beams in each tiling, the shape of the tiling boundary, their overlap ratio or the boundary size and the central position of the tiling. As noted above, for an optimal gain response the beams should be tiled around the array boresight where the directional gain of the antennas is maximised. However, for more flexibility, \textsc{Mosaic} supports the generation of an arbitrary number of off-boresight tilings to support observations in which there are multiple extended sources of interest within the telescope FoV. With the requested tiling configurations, \textsc{Mosaic} derives the corresponding semi-axes and the orientation angle from the beam model and generates a set of beam positions in equatorial coordinates. At this stage, the observer may also specify independently tracking beams not in a tiling to cover known point sources within the FoV. 
  \item \textbf{Delay polynomial generation:} For each of the beam coordinates $b_i$ generated in the preceding step, the geometric delay terms $d_{b_i}(t)$, are calculated. Additionally, the geometric delay to the array phase centre, $d_{\circ}(t)$, is calculated. As the data received by FBFUSE have already undergone delay correction to the array phase centre, it is the differential delay between each beam position and array phase centre, $\delta d_{b_i}(t) = d_{b_i}(t) - d_{\circ}(t)$, that is required by FBFUSE to produce coherent beams. The delay values, $\delta d_{b_i}(t)$, are calculated for each beam and a first order polynomial is fit to the values. FBFUSE periodically invokes \textsc{Mosaic} to calculate new polynomials which are evaluated inside the beamformer to produce and update weights. The minimal weight update interval (MWUI) is defined as the length of the time before the position error of the beam caused by the rotation of the earth, becomes as large as 1\% of the diameter of the beam (corresponding to a 0.028\% loss of sensitivity for a Gaussian beam). The length of MWUI is shorter if the distance between the beam and the boresight is larger and the size of the beam is smaller. Therefore, MWUI is calculated considering the beam is formed at the edge of the primary beam with the longest baseline (7697.58 meters). The value of MWUI is 481 ms for the abovementioned situation, while in a practical MeerKAT observation, 100 ms is used to compensate for the change of the beam shape during the interval. In the meantime, the delay polynomial is produced in every 2 seconds. 
\end{itemize}
 
\subsection{Point source survey speed}
The point-source survey speed achievable with beamformed observations with MeerKAT is dependent on the array configuration, sensitivity of the individual antennas, beamforming efficiency and the total number of beams that can be produced. We estimate the survey speed figure of merit (\textrm{FoMSS}) using a modified version of equation (3) of \citet{Cordes_SKA_Memo2009}:
\begin{equation}
FoMSS = B\frac{\Omega_{\rm{beam}}}{N_{\rm sa}}\left(\frac{f_cA_e}{mT_{\rm{sys}}}\right)^2 \sum_{i=1}^{N_{\rm beam}} g_{i}^{2},
\label{eq:FOMSS}
\end{equation}
where $B$ is the bandwidth of the receiver, $N_{\rm{beam}}$ is the number of synthesized beams, $\Omega_{\rm{beam}}$ is the field of view of a single synthesized beam, $N_{\rm sa}$ is the number of sub arrays, $A_e$ is the effective collecting area of the array, $f_c$ is the fraction of $A_e$ that is available, $m$ is the threshold of signal-to-noise ratio (S/N) for detection, $T_{\rm{sys}}$ is the system temperature and $g_i$ is the fraction of the peak of the primary beam gain at the position of synthesised beam $i$. The model of the primary beam was obtained using the \textsc{katbeam}\footnote{\url{https://github.com/ska-sa/katbeam}} package which uses a cosine-tapered field model as suggested by \citet{Mauch2020}. The gain within a coherent beam was considered constant in this calculation. Currently, the FBFUSE beamformer may only be used in a single MeerKAT subarray making $N_{\rm sa}=1$. \par
\begin{figure*}[h]
\begin{center}
  \includegraphics[width=0.49\textwidth]{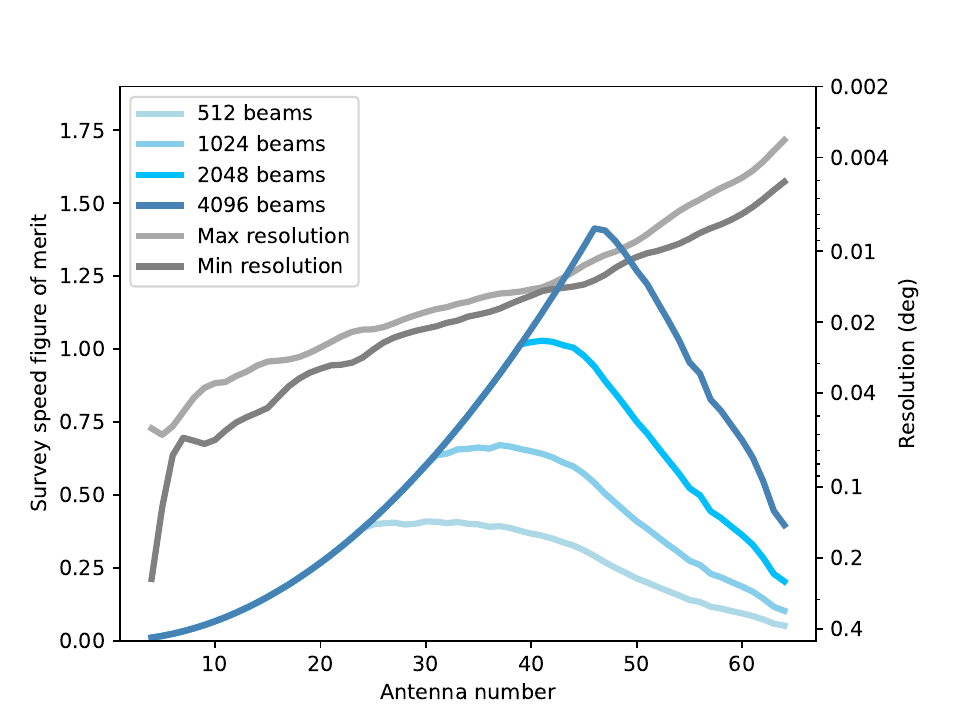}
  \includegraphics[width=0.49\textwidth]{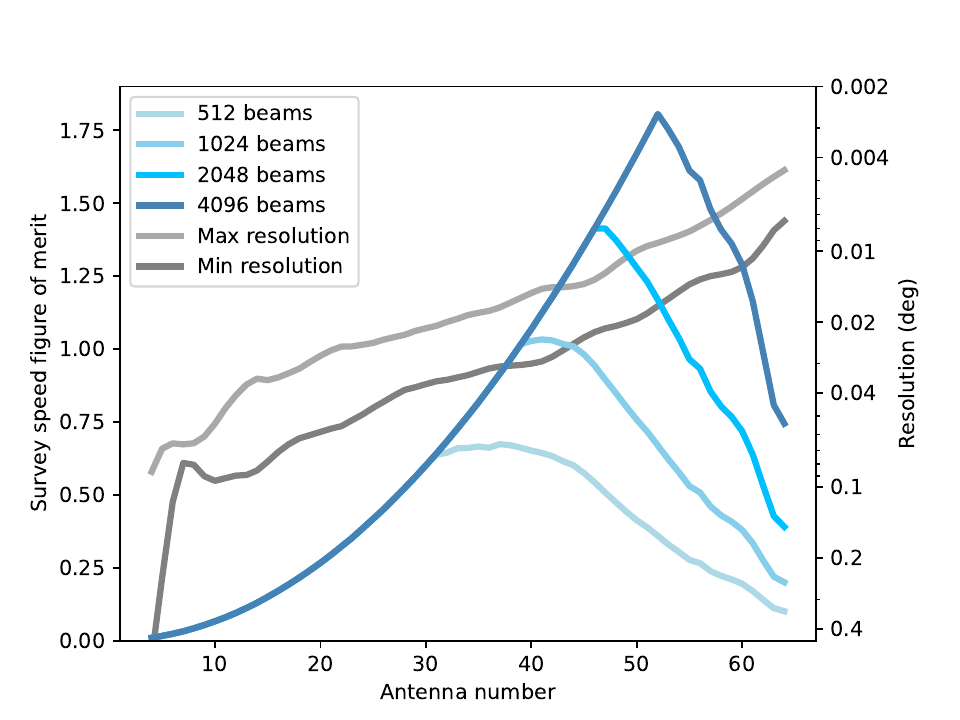}
\end{center}
\caption{Survey speed figure of merit ($\rm{FoMSS}$) as a function of array size when the source is at zenith (left) and at an elevation of 30$^{\circ}$ (right). Antennas are included in the array in order of increasing the distance from the array centre. The $\rm{FoMSS}$ is shown for four different maximum beam numbers (we note that although the current FBFUSE beamformer can only produce $\sim1000$ beams for the full array, it can produce more for smaller arrays due to the lower computational cost of beamforming with fewer antennas). The maximum and minimum angular resolution of the simulated beam for each array size is shown in gray.}
\label{fig:surveySpeed_vs_antNum}
\end{figure*}

Figure \ref{fig:surveySpeed_vs_antNum} shows the $\rm{FoMSS}$ for multi-beam observations with MeerKAT under different array configurations, numbers of beams and source elevations. Here we consider a simple array definition based on the incremental inclusion of dishes in order of increasing distance from the array centre. Initially, the $\rm{FoMSS}$ values increase because of the addition of the core dishes and the integrated FoV of the beams exceeding the FoV of the primary beam (here defined as the FWHM of the primary beam at an arbitrary frequency). However, as more dishes further away from the core are included, the beams become smaller and eventually reach a turning point where they can no longer fill the primary FoV. The array size that maximises the $\rm{FoMSS}$ for different numbers of beams and source elevations is provided in Table \ref{tab:suvery_speed_turning_point}. For a typical observation using approximately 1000 beams, we find that the optimal number of antennas to use for survey observations is between 37 and 41 depending on the average elevation of observations.

\begin{table}[h!]
    \caption{The array size (in number of antennas included radially from the array centre) required to maximise the survey speed figure of merit ($\rm{FoMSS}$) for different numbers of synthesised beams.}
    \begin{center}
    \begin{tabular}{p{1cm} | p{1cm} p{1cm} p{1cm} p{1cm}}
         \hline
         \hline
         & \multicolumn{4}{c}{Elevation}\\
         \hline
           $N_{\rm{beams}}$ & 90$^{\circ}$ & 60$^{\circ}$ & 45$^{\circ}$ & 30$^{\circ}$ \\ \hline
            512 & 33 & 33 & 33 & 37 \\
            1024 & 37 & 37 & 38 & 41 \\
            2048 & 41 & 42 & 44 & 47 \\
            4096 & 46 & 48 & 49 & 52 \\
        \hline
    \end{tabular}
    \end{center}
    \label{tab:suvery_speed_turning_point}
\end{table}

\subsection{Tiling validity}
\label{sec:tiling_validity}
\begin{figure*}[h!]
  \includegraphics[width=0.32\textwidth]{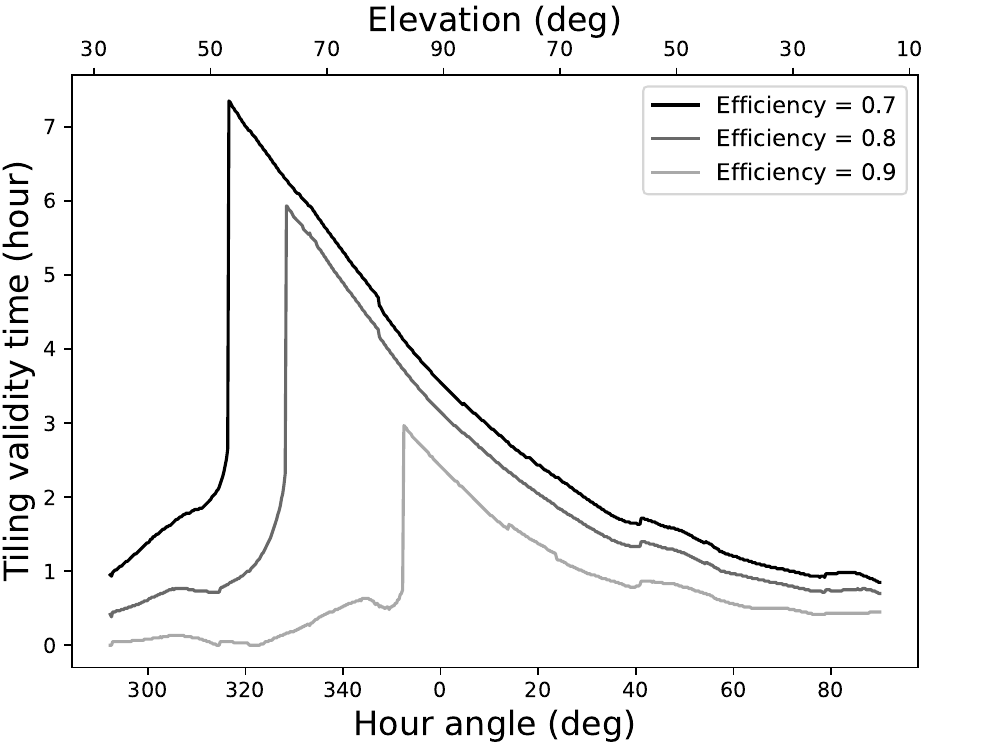}
  \includegraphics[width=0.32\textwidth]{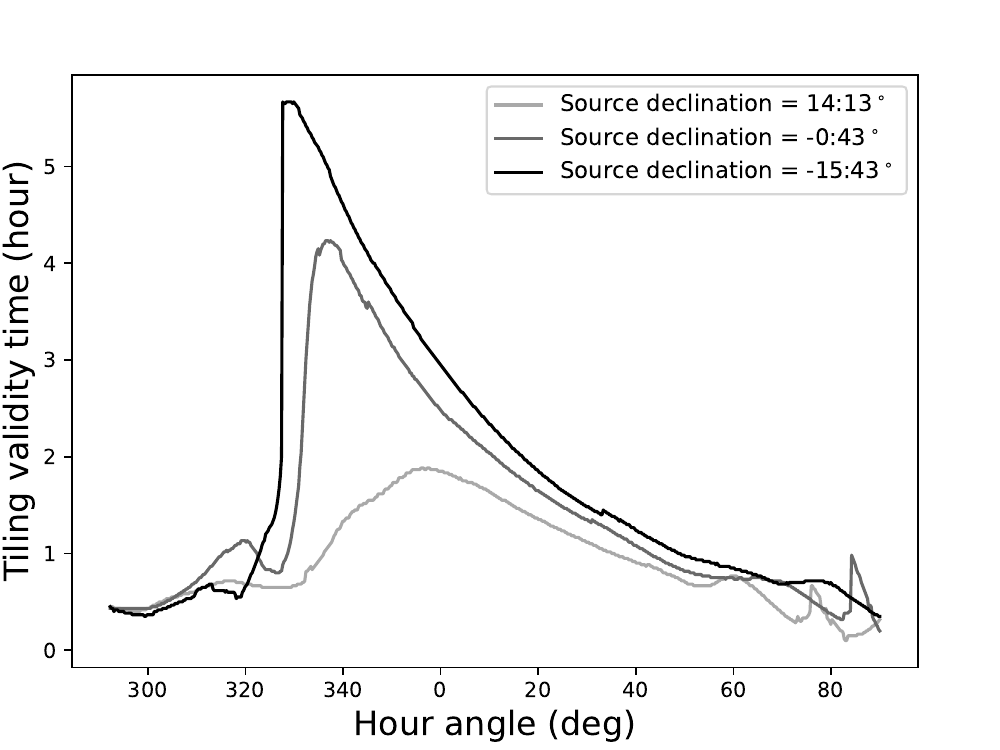}
  \includegraphics[width=0.32\textwidth]{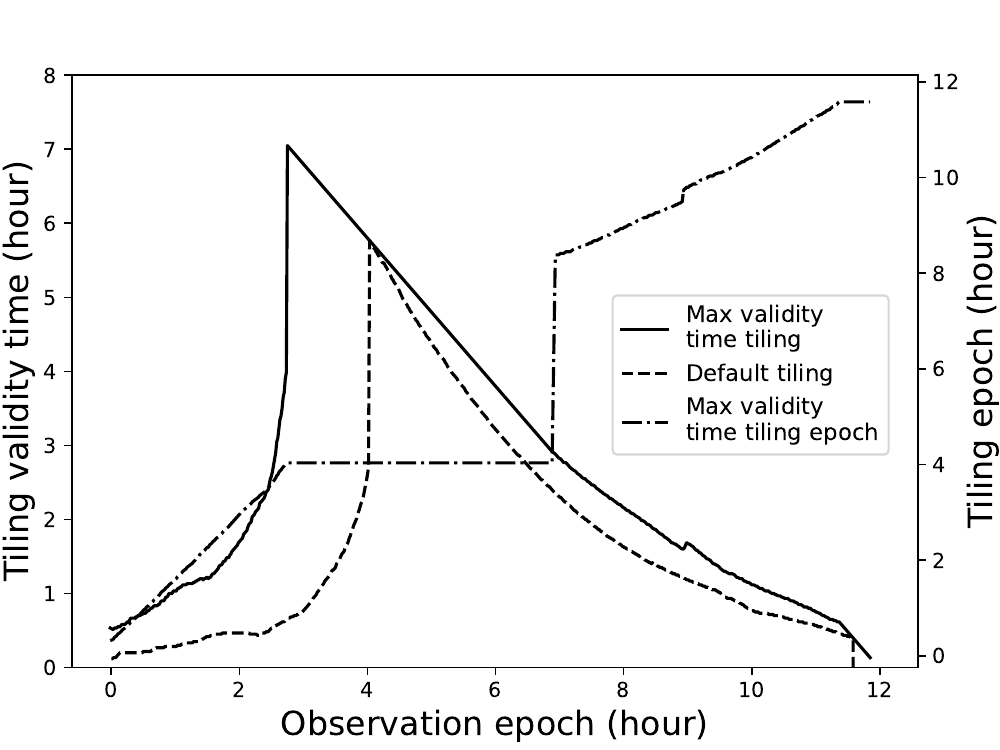}
   \caption{Left: The tiling validity time as a function of starting hour angle at various efficiency thresholds for a source that transits through the zenith at MeerKAT. Middle: The tiling validity time for an efficiency 0.8 tiling as a function of starting hour angle for sources at a range of declinations. Right: The maximum validity time as a function of observation epoch for a source transiting through the zenith with an efficiency 0.8 tiling. The solid line is the max validity time that can be achieved when the observation starts with a non-optimal tiling, while the dashed line is the result when starting with an optimal tiling. The dashed-pointed line is the epochs of the non-optimal tiling for difference observation epoch to achieve max validity time (see the last paragraph of section \ref{sec:max_validity_time} for an example).}
   \label{fig:validity_time_of_tiling}
\end{figure*}

For long duration observations, it may be necessary to re-tile the beams as the current tiling becomes less efficient due to the changes in the beam shape as illustrated in Figure \ref{fig:overlap_evolution}. It is therefore useful to determine how long before a tiling generated at the start of that observation drops below some pre-defined efficiency. Here we simulated the beam shape evolution with time for the full MeerKAT array to determine the validity time of a tiling at a given declination. \par
Simulated observations start at timestamp $t_0$ with an optimized tiling. The beam shape is then updated with an interval $\delta t$ (1 minute). At each update, a new beam shape is generated for that epoch and the resultant beams are positioned at the tiling coordinates from $t_0$. The tiling efficiency (defined in Section \ref{sec:tiling_evolution}) is then measured. If it exceeds the desired efficiency, the simulation will continue. If the target efficiency is not met anymore, the simulation is stopped and the time since $t_0$ is recorded as the tiling validity time. The process then repeats with a new simulation starting at time $t_1 = t_0 + \delta t$. This process is continued until the source drops below 10$^{\circ}$ in elevation. The simulated validity time with different efficiency thresholds is plotted in the left panel of Figure \ref{fig:validity_time_of_tiling}. This shows the validity time of the tiling as a function of the hour angle for a source that transits through the zenith at MeerKAT. The simulation shows that the validity time has a sharp peak before the source reaches the zenith, then gradually decreases for the rest of the simulation, with the peak of the validity time occurring earlier for lower efficiency thresholds. \par
The middle panel of Figure \ref{fig:validity_time_of_tiling} shows the tiling validity for a range of source declinations. Here we see that declinations that result in lower-elevation source transits have shorter validity times. This can be understood by considering that at lower elevations the beams have a higher degree of asymmetry and as such the rotation of these beams will have a larger effect on the tiling efficiency. \par 
If we start an observation in which the tiling is initially non-optimal but transitions into an optimal configuration with time, the ``max validity time'' should be longer than the scenario in which the observation starts with an optimal tiling. This can be seen by simulating the max validity time using the same source of the left panel of Figure \ref{fig:validity_time_of_tiling} with a tiling efficiency of 0.8. The virtual observation started at 6 am with a duration of 12 hours. The duration was divided into 720 timestamps. The simulation started at timestamp $t_0$ and calculated a corresponding beam shape, based on which a tiling $c_{t_0}$ was also generated. Then the simulation moved to $t_{-1}$ which was the closest timestamp before $t_0$ (if existing). The beam shape of timestamp $t_{-1}$ was calculated and it was put into tiling $c_{t_0}$ for the evaluation of the tiling efficiency. If the tiling efficiency was above the threshold, the simulation would continue to move to timestamp $t_{-n}$ where $n = 2, 3, 4, ...$, otherwise, the simulation would move to timestamp $t_{+1}$. Similarly, a new beam shape was generated and put into the tiling $c_{t_0}$ for efficiency evaluation. If the threshold is not reached, the simulation would move to $t_{+n}$, otherwise, the optimal validity time for timestamp $t_0$ is $t_{+n-1} - t_{-n+1}$. In this manner, the simulation iterated all timestamps and the result is plotted in the right panel of Figure \ref{fig:validity_time_of_tiling}. The solid line in the plot is the max validity time starts with a non-optimal tiling against the timestamps, while the dashed line is the validity time starts with an optimal tiling. As can be seen from the comparison, the former approach produces a longer validity time than the latter and moves the sharp jump earlier. Additionally, the dash-dotted line indicates the best tiling for a given timestamp. For example, the best tiling to achieve a long validity time for timestamp 8-hour is the tiling generated from the beam shape near timestamp 9-hour. The straight line from 2.7-hour to 6.9-hour indicates that the tiling generated at the beam shape of the 4-hour timestamp produces the overall best validity time for that period. \label{sec:max_validity_time}

\subsection{A beamformed observation of 47 Tucanae}
We carried out two observations using the MeerKAT MPIfR multi-beam system to test and verify the beamforming and tiling system.
\subsubsection{MeerKAT observations}
At 06:00 UTC on May 2, 2020, a 4-hour observation of the globular cluster 47 Tucanae (NGC 104) was made. A total of 56 antennas were used with the majority being in the MeerKAT core. The observing band was centered at 1.284 GHz with a bandwidth of 865 MHz which was divided into 256 channels. During the observation, the source went from an hour angle of -33.4$^\circ$ to 26.8$^\circ$. \textsc{Mosaic} was invoked to produce a tiling of 264 beams with an overlap ratio of 0.7. The tiling covered a region of approx. 0.45$^{\circ}$ in R.A. and 0.14$^{\circ}$ in declination. Additionally there were 24 tracking beams constantly pointing at 24 of the known pulsars in the cluster. The packing of the tiling on the cluster is plotted in the central panel of Figure \ref{fig:profile_overlay} overlying an optical image of 47 Tucanae. Another 440-s observation on the pulsar PSR~J1644-4559 was carried out at 11:00 UTC on Dec 1, 2020 with 54 antennas and an overlap ratio of 0.996 to verify the beam shape.

\subsubsection{Beamforming Efficiency}
\begin{figure*}[ht]
  \includegraphics[width=1\textwidth]{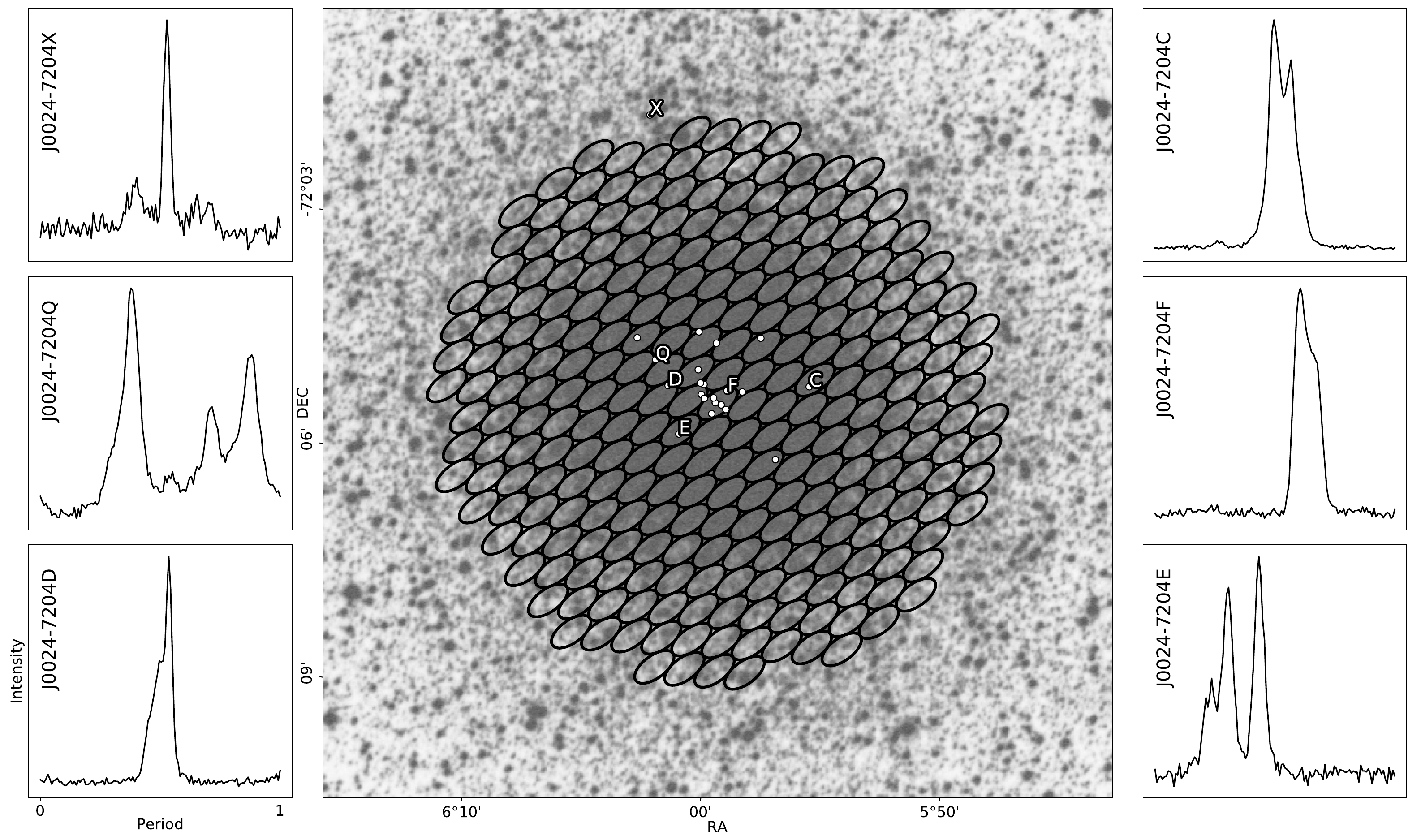}
   \caption{Profiles of selected pulsars and the tiling pattern used for observation of 47 Tucanae. In the central panel, the background is an optical image of 47 Tucanae obtained from the STScI Digitized Sky Survey \citep{Lasker1994}, the ellipses are the tiling of the coherent beams at the start of the observation and the white dots indicate the positions of the known pulsars in the cluster. The average profiles in the left and right panels are folded from the tracking beams on the pulsars whose locations are indicated by their letter designation in the central panel.}
   \label{fig:profile_overlay}
\end{figure*}

The first test was to check the efficiency of the beamforming process as it is the foundation of other applications. If the weight calculation during the beamforming process is incorrect or not accurate enough, the signal from the known pulsars might not be detectable because the antenna data are combined out of phase. To assess this situation, we folded the data of the tracking beams using the corresponding ephemerides of each pulsar. The average profiles for six of these pulsars are plotted in the left and right panels of Figure \ref{fig:profile_overlay} with their positions labeled in the central panel. The average profiles shown are a close match to those obtained by \citet{Freire2000} and \citet{Ridolfi2017}. \par

To determine the efficiency of the beamforming process, we compared the signal-to-noise ratio (S/N) between the incoherent beam and the coherent beams. For the phased addition of $N$ voltages with uncorrelated noise, the signal power increases as $N^2$ while the noise power increases as $N$. As such, the S/N of the observed pulsars in the coherent beam should scale as $N$. For the incoherent beam, the power in each antenna is first detected and then summed. Here the r.m.s. of the noise increases as $\sqrt N$ and as such the S/N of the observed pulsars in the incoherent beam should scale as $\sqrt N$. The ratio of S/Ns between the coherent to incoherent beam should thus be $N / \sqrt{N} = \sqrt{N}$. Note that the primary beam gain scaling is the same for both the incoherent and coherent beams for a given position in the primary beam and as such the position of the pulsars in the FoV has no impact on the coherent to incoherent S/N ratio.
For an ideal beamformer, the expected gain from the incoherent beam to the coherent beam for this observation with 56 antennas should be $\sqrt{56} = 7.48$. \par
We folded the incoherent beams and the coherent beams of the pulsar J0024$-$7204C, J0024$-$7204D and J0024$-$7204E using \textsc{DSPSR}\footnote{\url{http://dspsr.sourceforge.net}} and compared their S/N listed in Table \ref{tab:SNR_between_incoherent_coherent_beam}. The S/N were measured using the \texttt{pdmp} tool in the \textsc{PSRCHIVE}\footnote{\url{http://psrchive.sourceforge.net}} software package. It estimates and subtracts the baseline of the data, then uses a matched filter to determine the width of the pulse and calculate the S/N. To account for imperfect folding ephemerides, \texttt{pdmp} optimizes over both period and dispersion measure (the integrated free electrons column density between the observer and the source). The efficiency of the beamforming process is then calculated as:  
\begin{equation}
\mathrm{Efficiency} = \frac{1}{\sqrt{N} }\frac{{\rm S/N}_{\mathrm{coherent}}}{{\rm S/N}_{\mathrm{incoherent}}}
\label{eq:beamforming_efficiency}
\end{equation}
\begin{table}[h!]
\caption{The S/N of pulsars C, D and F in the coherent and incoherent beams and the derived beamforming efficiency. The efficiency is calculated using Equation \ref{eq:beamforming_efficiency}. We note that the incoherent beam is more affected by RFI than the coherent beams which can lead to boosted beamforming efficiencies when using this method.}
\begin{center}
\begin{tabular}{llll}
\hline
\hline
           & J0024$-$7204C & J0024$-$7204D & J0024$-$7204F \\ \hline
${\rm S/N}_{\mathrm{incoherent}}$ & 82       & 25       & 35       \\
${\rm S/N}_{\mathrm{coherent}}$   & 642      & 177      & 270      \\
Efficiency & 105\%    & 94\%     & 103\%       \\ \hline
\end{tabular}
\label{tab:SNR_between_incoherent_coherent_beam}
\end{center}
\end{table}

The incoherent beam is more susceptible to contamination by RFI and variations in the system noise of individual antennas. For a coherent beam, these effects are partially mitigated as they sum out of phase during the coherent beamforming (so-called ``phase-washing''). This effect can lead to the efficiency of the beamforming being overestimated. The high efficiencies seen here demonstrate that the calculation of the weights is accurate enough to achieve high coherency.   

\subsubsection{Verification of the simulated beam shape}
\label{sec:beamshape_verification}
\begin{figure*}[h!]
  \includegraphics[width=0.48\textwidth]{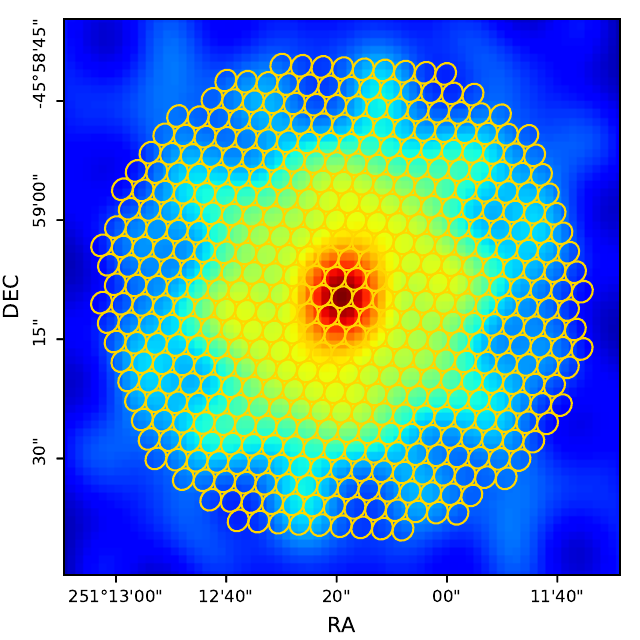}
  \includegraphics[width=0.48\textwidth]{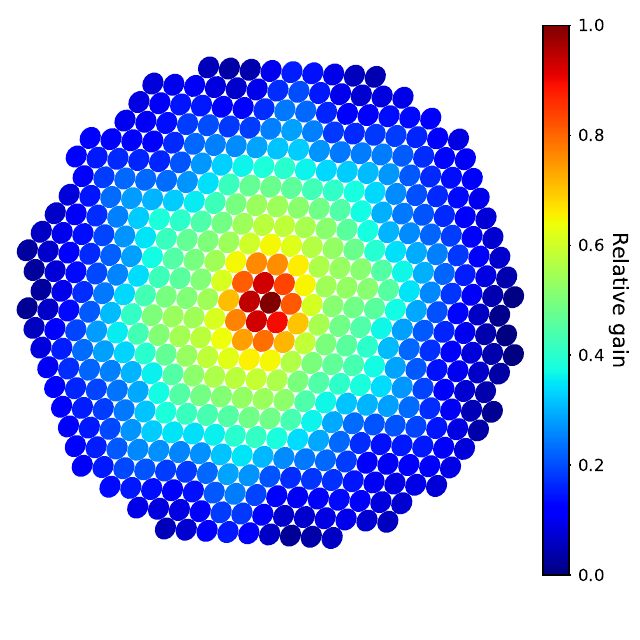}
   \caption{Verification of the beam shape: The image on the left is the simulated PSF covered by a tiling. An overlap ratio of 0.996 was used for the purpose of supersampling the PSF. The image on the right is the power distribution of the sky based on a real observation beamformed using the same tiling, array configuration and boresight pointing. The ellipses represent the beams and the color is normalized according to their S/N with the colour bar representing the relative gain of the beam.}
   \label{fig:beamshape_verfification} 
\end{figure*}

The second test was the verification of the simulated beam shape by comparing it to the power distribution of the sky derived from the observation data of PSR~J1644$-$4559. First, we simulated the beam shape of the array pointed to the source at the planned observation time. Then we observed the source with a very high overlap ratio tiling to densely sample the PSF as illustrated in the left panel of Figure \ref{fig:beamshape_verfification}. The beamformed data were folded and analysed to obtain the S/N measurements of the pulsar in each beam. The observed beams are represented by ellipses with colors normalized by their S/N and plotted in the right panel of Figure \ref{fig:beamshape_verfification} according to their sky position. The comparison shows the power distribution of the sky is in high agreement with the simulated PSF including the complex structure at the outer region of the beam shape. This also further confirms the correctness of the weighting calculations and the tiling generation. We note that there is a small position offset of $\sim$0.8 arcsecond in the beamformed data probably due to a minor phasing error. 

\subsubsection{Tracking the change of the beam shape and the tiling}
The third test performed was the verification of the change of the beam shape and the tiling by tracking variations in S/N of 47 Tucanae pulsars in multiple beams. There was no re-tiling during the four-hour observation, so the efficiency of the tiling was expected to vary due to the changes of the beam shape as a function of time as discussed in Section \ref{sec:tiling_evolution} and illustrated in Figure \ref{fig:overlap_evolution}. As a result, the relative distances between a pulsar and the surrounding beams also change which manifests in their S/N. By simulating the evolution of the tiling, we can predict the trend of the change of the S/N of those beams and verify them with the data.\par
\begin{figure*}
\centering
\includegraphics[width=0.88\textwidth]{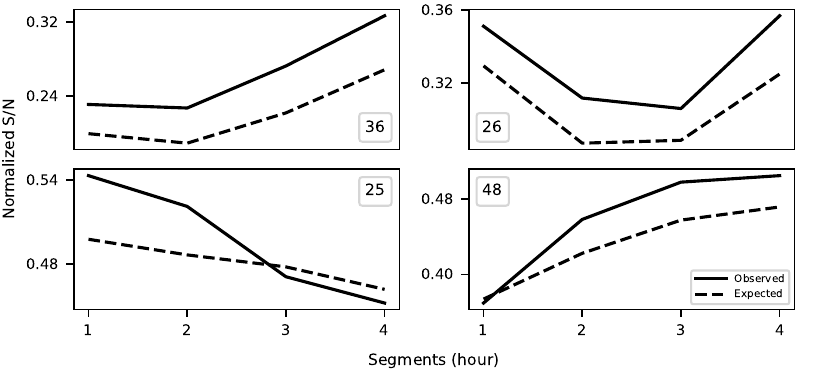}
\caption{The change of the observed S/N (solid line) and expected value (dashed line) for the beams surrounding the position of PSR~J0024$-$7204F, i.e. beams 25, 26, 36 and 48. The corresponding beams are noted in each panel.}
\label{fig:surrounding_beams_SNR_trend}
\end{figure*}

To perform this test, the observation was divided into four, one-hour segments and the surrounding beams of each pulsar were folded using the known ephemeris. To mitigate temporal variations due to interstellar scintillation, the S/N observed in the surrounding beams was normalised by the S/N in the corresponding tracking beam pointing directly to the respective pulsar. To reduce the effect of the changing PSF over the wide observing bandwidth, only the band between 1.4--1.5 GHz was used. The positions of the pulsar in the PSF of each of the surrounding beams were used to predict their respective S/Ns. The results for the beams surrounding PSR~J0024$-$7204F (beams 25, 26, 36 and 48) are plotted in Figure \ref{fig:surrounding_beams_SNR_trend}. To account for the changing parallactic angle, the average PSF over each one-hour segment was used. In all cases the PSFs were generated at a reference frequency of 1.45 GHz.

\par
The result from the observation is in overall agreement with the expected tiling evolution. Though the trend of the S/N mostly follows the expected evolution, there is apparently an offset in amplitude between the two. This is likely due to the fact that we do not account for uncertainties in our S/N estimation, scintillation of the source due to propagation through the interstellar medium, and imperfect antenna weighting which leads to differences between the modelled and actual synthesised beam of the array. Despite the offsets in amplitude, the overall agreement between the expectation and the observation indicates that the tiling is correctly generated and implemented in the sky. It also suggests that the simulated PSF can largely represent the telescope gain on the sky and provides a verification of the evolutionary behaviour of the tiling shown in Figure \ref{fig:overlap_evolution}.

\subsubsection{Localization of a point source using multiple beams}
\label{sec:multi_beam_localization}
With the tiling verified, the position of a pulsar can be estimated using the S/N of its detections in multiple beams. As a demonstration of this capability, we have estimated the position of a pulsar in 47 Tucanae using a Markov Chain Monte Carlo method and compared them with the known pulsar position. To fit the position, we first assume that the pulsar is at an arbitrary position $x$ inside multiple beams $N_{i}$, where $i$ is the index of the beam. If the S/N of a beam centered on the pulsar is $R_P$, then in an ideal situation, the S/N of the beam $N_{i}$ would be $R_{N_{i}} = R_P \times P_{N_{i}}(x)$ where $P_{N_{i}}(x)$ is the sensitivity of the beam $N_{i}$ at position $x$. The values of $R_{N_{i}}$ can be measured on the observational data and the function $P_{N_{i}}(x)$ can be constructed with a model of the sensitivity of the beam, e.g. the PSF. With those, the fitting algorithm generates samples of $x$ and evaluates them for multiple beams. A similar method has been proposed by \citet{Obrocka2015} which uses multiple tied-array beams and spectral information to estimate the position of a transient source. 
We fitted the positions of PSR~J0024$-$7204F with this method. The localization results and their 95\% credible intervals are listed in Table \ref{tab:localization_difference}. The known position of the pulsar is 00:24:03.8554(1), -72:04:42.8183(1), which is taken from the ATNF pulsar catalog\footnote{\url{https://www.atnf.csiro.au/research/pulsar/psrcat/}} \citep[\texttt{psrcat},][]{manchester2005}

\begin{table}[h]
\caption{The localizations of PSR~J0024$-$7204F in each of the corresponding four, hour-long segments with their 95\% credible intervals.}
\centerline{
\begin{tabular}{lll}
\hline \hline
Segment      &  RA   &  DEC   \\ \hline
1            & 00:24:03.8$^{+1.7}_{-1.7}$ & -72:04:44.6$^{+8.6}_{-7.6}$  \\
2            & 00:24:03.7$^{+1.8}_{-1.8}$ & -72:04:43.1$^{+10.0}_{-8.9}$ \\
3            & 00:24:03.7$^{+1.8}_{-1.8}$ & -72:04:42.6$^{+11.3}_{-9.0}$ \\
4            & 00:24:03.8$^{+1.7}_{-1.7}$ & -72:04:43.9$^{+9.9}_{-7.5}$ \\ \hline
\end{tabular}
}
\label{tab:localization_difference}
\end{table}

The localization of the pulsar with multiple beams is largely good. Again, here the observed offsets are most likely due to the aforementioned limitations in our ability to accurately model the PSF. To mitigate the effects of interstellar scintillation, we limit the bandwidth used for localization to 100 MHz. A reduced bandwidth leads to a decrease in S/N which affects the quality of the localization. Nevertheless, these results show that the localization of any detection can be improved (within the size of the beam). In general, better S/N produces better localization. It is also beneficial if the true position of the source lies inside the region where the absolute derivative of the PSF is large, as the predicted S/N changes rapidly there, leading to smaller uncertainties in the estimated localization. A more complete approach would be to use a time and frequency dependent PSF modelling that accounts for the source spectrum and scintillation. This will be addressed in the forthcoming Bezuidenhout et al. in prep.

\section{Summary and Outlook}
\label{sec:summary}
In the first part of the paper, we described a general beamforming approach applicable to modern interferometers with a large number of receiving elements. This enables wide field, high time and spatial resolution observations with the sensitivity of all elements coherently combined. To explore how multi-beam observations using this approach can be conducted, a software package called \textsc{Mosaic} has been developed. The package provides tools to simulate and model the beam shape, generate optimized tiling patterns, and predict the evolution of the tiling through time with the choice of performing a re-tile automatically when specific criteria are met.\par   
In the second part of the paper, we discussed practical applications for the aforementioned techniques by applying them to the MeerKAT telescope. It described how beamforming is implemented on MeerKAT and its use during observations. To demonstrate the planning of a multi-beam observation, the validity times of the tilings in different array configurations and source positions are provided. \par 
Finally, we provide the result of real observations using the MPIfR-developed FBFUSE beamformer. These demonstrate the capability of the system to observe multiple pulsars simultaneously with the expected sensitivity and show a high level of agreement between the simulated beam shape and the reconstructed beam shape obtained from a high overlap ratio observation. The consistency between the predicted and observed S/N in each beam around the pulsar location, allowing for arcsecond position determination, demonstrates that the complete system from beamforming to PSF determination and tiling operates as expected for MeerKAT.\par   
The MeerKAT telescope has been using \textsc{Mosaic} and FBFUSE to carry out multi-beam science observations since 2019. Three principal science programs rely on the systems presented in this paper. The Transients and pulsars with MeerKAT (TRAPUM) survey is a pulsar search targeting globular clusters, unassociated FERMI LAT sources, nearby galaxies and miscellaneous high-energy sources. The TRAPUM project seeks to discover millisecond and binary pulsar systems that can be used to answer questions about fundamental physics, the pulsar emission mechanism and stellar evolution. The survey uses various configurations depending on the exact sources and science goals of each different target class. Typically TRAPUM observations use between 277 and 480 beams depending on the observing frequency used. For observations requiring the widest FoV, the MeerKAT core is used, while for observations of targets with small extents (e.g. globular clusters) the full array is used. Currently, TRAPUM operates at L-band (856 -- 1712 MHz) and U-band (544 -- 1088 MHz) with plans to include S-band observations when the receivers become available \citep{Kramer2016}.\par
Complementing the TRAPUM project is the 3000-hour MPIfR Galactic Plane Survey (MGPS). This project consists of three parts: a shallow L-band Galactic plane survey, a deep, but limited latitude range, an S-band Galactic plane survey and an ultra-deep 200-hour survey of the region around Sgr A*. The MGPS survey seeks to greatly increase the number of known pulsars in the Galactic field with an emphasis on the discovery of relativistic binaries. Additionally, the Galactic centre portion of the search has the strongest chance to detect a pulsar in orbit around Sgr A*. The MGPS survey uses the MeerKAT core with 480 beams (Padmanabh et al. in prep.) Discoveries from the TRAPUM and MPGS project are made public via the project webpage\footnote{\url{http://trapum.org/discoveries.html}}.
\par
More TRAnsients and Pulsars (MeerTRAP) is a project to search for fast transients commensally with other MeerKAT science programs. It performs real-time, GPU-based, time-domain searches for short duration transients such as FRBs and rotating radio transients (RRaTs). It is planned that upon the detection of a new interesting source, the MeerTRAP project will provide a trigger to the transient buffer implemented in the beamformer. This allows for robust localisation and polarisation analysis for discovered transients. A demonstration of the system has been made through observations of FRB 121102 \citep{Caleb2020}. Due to its nature as a commensal project, MeerTRAP has a diverse array of configurations. Typically it uses the MeerKAT core with 480 or 864 beams depending on the configuration on the MeerKAT F-engines. \par
The upcoming SKA-mid \citep{Swart2020} will be a similar array to MeerKAT with an increased number of antennas and longer baselines. Therefore, the projects mentioned above are demonstrating capabilities envisaged for SKA-mid but at a smaller scale. Furthermore, the challenges and solutions that we discussed in this paper or encounter in those projects could be applicable for this next generation interferometric array. 

\section{Acknowledgment}
We thank for the help from the commission group and committee of MeerKAT and the help from the TRAPUM group. We thank for the help from Benjamin Hugo to the development of the \textsc{Mosaic} and a useful conversation on localisation with Vivek Venkatraman Krishnan. We also thank Tim Sprenger for the idea on calculating the validity time of a tiling and Jason Wu for the help with the observations. We thank the STScI digitized sky survey produced at the Space Telescope Science Institute for the optical image of 47 Tucanae. The work of this paper and the development of $\textsc{Mosaic}$ was funded by BlackHoleCam, the financial support by the European Research Council for the ERC Synergy Grant BlackHoleCam (ERC-2013-SyG, Grant Agreement no. 610058) is gratefully acknowledged. The development of FBFUSE is funded by the Max-Planck-Gesellschaft. Benjamin W. Stappers acknowledges funding from the European Research Council (ERC) under the European Union's Horizon 2020 research and innovation programme (Grant Agreement No. 694745). The MeerKAT telescope is operated by the South African Radio Astronomy Observatory, which is a facility
of the National Research Foundation, an agency of the Department of Science and Innovation.
\bibliographystyle{elsarticle-harv} 
\bibliography{references}

\end{document}